\documentclass[a4paper]{jpconf}
\usepackage{graphicx}
\usepackage{subcaption}
\usepackage{soul}

\usepackage[utf8]{inputenc}          
\usepackage{graphicx}         
\usepackage{array,dcolumn,longtable} 
\usepackage{amsmath,amssymb,slashed} 

\usepackage[table]{xcolor}
\usepackage{color}
\usepackage{colortbl}
\usepackage{multirow}

\definecolor{darkred}{rgb}{0.8, 0.0, 0.0}
\definecolor{midred}{rgb}{0.94, 0.19, 0.22}
\definecolor{litered}{rgb}{0.91, 0.33, 0.5}
\definecolor{superlitered}{rgb}{0.99, 0.56, 0.67}
\definecolor{darkblue}{rgb}{0.0, 0.0, 0.55}
\definecolor{midblue}{rgb}{0.12, 0.56, 1.0}
\definecolor{liteblue}{rgb}{0.0, 0.72, 0.92}
\definecolor{superliteblue}{rgb}{0.49, 0.98, 1.0}
\definecolor{darkgreen}{rgb}{0.09, 0.45, 0.27}
\definecolor{midgreen}{rgb}{0.52, 0.73, 0.4}

\begin{document}
\title{Far From Equilibrium Hydrodynamics and the Beam Energy Scan}

\author{Travis Dore}
\address{University of Illinois at Urbana-Champaign, Urbana, IL 61801, USA}
\author{Emma McLaughlin}
\address{Department of Physics, Columbia University, 538 West 120th Street, New York, NY 10027, USA}
\author{Jacquelyn Noronha-Hostler}
\address{University of Illinois at Urbana-Champaign, Urbana, IL 61801, USA}

\begin{abstract}
The existence of hydrodynamic attractors in rapidly expanding relativistic systems has shed light on the success of relativistic hydrodynamics in describing heavy-ion collisions at zero chemical potential.  As the search for the QCD critical point continues, it is important to investigate how out of equilibrium effects influence  the trajectories on the QCD phase diagram. In this proceedings, we study a Bjorken expanding hydrodynamic system based on DMNR equations of motion with initial out of equilibrium effects and finite chemical potential in a system with a critical point.  We find that the initial conditions are not unique for a specific freeze-out point, but rather the system can evolve to the same final state freeze-out point with a wide range of initial baryon chemical potential, $\mu_B$.  For the same initial energy density and baryon density, depending on how far out of equilibrium the system begins, the initial $\mu_B$  can vary by $\Delta \mu_B\sim 350$ MeV. Our results indicate that knowledge of the out-of-equilibrium effects in the initial state provide vital information that influences the search for the QCD critical point.
\end{abstract}

\section{Introduction}
Determining the initial conditions for heavy-ion collisions is an ongoing problem in the field, and no less so as the Beam Energy Scan program at RHIC continues its search for the QCD critical point \cite{Bzdak:2019pkr}. Studies of hydrodynamic attractors \cite{Heller:2015dha,Buchel:2016cbj,Heller:2016rtz,Spalinski:2017mel,Romatschke:2017acs,Romatschke:2017vte,Romatschke:2017acs,Behtash:2017wqg,Strickland:2017kux,Denicol:2017lxn,Blaizot:2017ucy,Casalderrey-Solana:2017zyh,Heller:2018qvh,Rougemont:2018ivt,Denicol:2018pak,Almaalol:2018ynx,Casalderrey-Solana:2018uag,Behtash:2018moe,Behtash:2019txb,Strickland:2018ayk,Strickland:2019hff,Kurkela:2019set,Jaiswal:2019cju,Denicol:2019lio} provide evidence that the effectiveness of hydrodynamics in describing heavy-ion data may come from the fact that the system quickly forgets about the initial conditions and collapses on to a universal attractor on short time scales. At LHC energies, where $\mu_B \sim 0$, the story is simpler as one does not need to worry about a multi-dimensional phase diagram. However, at lower beam energies where baryon stopping is relevant, as the system evolves it will traverse a path in the $T-\mu_B$ plane. If the system were to evolve entirely in equilibrium, it would simply follow isentropes (entropy over baryon number would be conserved i.e. $S/N_B=const$) in the phase diagram (see a number of recent studies relying on isentropes \cite{Gunther:2017sxn,Bellwied:2018tkc,Parotto:2018pwx,Noronha-Hostler:2019ayj,Monnai:2019hkn,Stafford:2019yuy}). However, the system evolves out of equilibrium, and depending on how far the system begins away from equilibrium, one would expect the deviations from the isentropic trajectories to be large.

Here, we study the effects of initializing the full $T^{\mu\nu}$, specifically the initial shear stress tensor $\pi^{\mu\nu}$ and initial bulk pressure $\Pi$, in an out of equilibrium state on the trajectories through the phase diagram. Previous studies of the full $T^{\mu\nu}$ on collective flow observables were only performed only at $\mu_B=0$ \cite{Liu:2015nwa,Kurkela:2018wud,Schenke:2019pmk} and, in this context, new causality conditions derived in \cite{Bemfica:2020xym} will be useful to constrain the far from equilibrium behavior. Here we run different initial conditions covering a wide range of initial shear and bulk viscous effects with a non-conformal equation of state that has a parametrized critical point. We find that far-from-equilibrium initial conditions can lead to significant deviations from isentropes, which implies connecting the initial state and final state can be significantly complicated at the beam energy scan. We also study whether or not hydrodynamic attractors persist in the presence of shear and bulk viscous terms that depend on both temperature and baryon chemical potential. 

\section{Hydrodynamic Model}
Relativistic viscous hydrodynamics has enjoyed much success in predicting many observables in the low transverse momentum sector of heavy-ion collisions. As the Beam Energy Scan program at RHIC continues, a number of new developments concerning the hydrodynamic modeling of heavy-ion collisions have been performed to study a baryon-rich environment \cite{Du:2019obx,Denicol:2018wdp,Batyuk:2017sku,Fotakis:2019nbq} and transport coefficients that are also dependent on $\left\{T,\mu_B\right\}$ \cite{Demir:2008tr,Denicol:2013nua,Kadam:2014cua,Monnai:2016kud,Rougemont:2017tlu,Rougemont:2017tlu,Auvinen:2017fjw,Martinez:2019bsn}. However, in order to find the critical point in the QCD phase diagram (if it exists) it will be necessary to have a thorough understanding of how a dynamical evolution past the critical point can influence final state observables \cite{Son:2004iv}. Moreover, an understanding of how out of equilibrium effects may influence criticality as well as system dynamics given two thermodynamic degrees of freedom, $T$ and $\mu_B$, is essential. 
\par
For the purpose of this proceedings, we will discuss only results obtained using DNMR hydrodynamic equations of motion \cite{Denicol:2012cn}. As a starting point for future work, we use a simplified Bjorken expanding model \cite{Bjorken:1982qr} to obtain qualitative results. To our knowledge, this is the first work to study out of equilibrium effects on trajectories in the phase diagram, especially in regards to their impact on the search for the critical point. To that end, a simplified Bjorken picture is ideal for obtaining a basic understanding of the sought after effects. For a Bjorken expanding fluid (where $\tau = \sqrt{t^2-z^2}$ is the propertime) the DNMR equations of motion become
\begin{eqnarray}
    \dot{\epsilon}&=&-\frac{1}{\tau}\left[e+p+\Pi-\pi^{\eta}_{\eta}\right]\\
      \tau_{\pi}\dot{\pi}^{\eta}_{\eta}+\pi^{\eta}_{\eta}&=&\frac{1}{\tau}\left[\frac{4\eta}{3} - \pi^\eta_\eta\left(\delta_{\pi\pi} + \tau_{\pi\pi} \right) + \lambda_{\pi\Pi}\Pi \right]\\
      \label{epDot}
 \tau_{\Pi}\dot{\Pi}+\Pi&=&-\frac{1}{\tau}\left(\zeta + \delta_{\Pi\Pi}\Pi+\frac{2}{3}\lambda_{\Pi\pi}\pi^{\eta}_{\eta}\right)\\
    \dot \rho &=& \frac{\rho_0}{\tau}\label{eqn:rhoBevo}
\end{eqnarray}
where $\epsilon$ is the energy density (with $\dot{\epsilon}= d\epsilon/d\tau$), $p$ is the equilibrium pressure, $\pi_\eta^\eta$ is the nonzero component of the shear stress tensor, $\Pi$ is the bulk pressure, $\rho$ is the baryon density ($\rho_0$ is the initial baryon density), $\eta$ is the shear viscosity, $\zeta$ is the bulk viscosity, and the remaining transport coefficients are given by \cite{Denicol:2018wdp,Bazow:2016yra}
\begin{eqnarray}
    \tau_\pi &=& \frac{5\ \eta}{\epsilon + p}\\
   \tau_{\Pi}&=&\frac{\zeta}{15(e+p)\left(\frac{1}{3}-c_s^2\right)^2}\label{eqn:tauPI}\\
    \lambda_{\pi \Pi} &=& \frac{6}{5} \tau_\pi\\
    \delta_{\pi\pi} &=& \frac{4}{3}\tau_\pi\\
    \tau_{\pi\pi} &=& \frac{10}{7}\tau_\pi\\
    \lambda_{\Pi \pi} &=& \frac{8}{5}\left(\frac{1}{3} - c_s^2\right)\tau_\Pi\\
    \delta_{\Pi \Pi} &=& \frac{2}{3}\tau_\Pi
\end{eqnarray}
where $c_s$ is the isentropic speed of sound. 
We use a temperature dependent shear viscosity ($\eta T/w$ where $w=e+p$ is the enthalpy), given by an excluded volume calculation below the transition temperature \cite{NoronhaHostler:2012ug} (using the PDG16+ \cite{Alba:2017mqu}) that is matched onto a QCD motivated parameterization in the deconfined phase \cite{Christiansen:2014ypa,Dubla:2018czx}.
\begin{figure}
    \centering
    \begin{subfigure}{.48\textwidth}
    \includegraphics[width=\linewidth,height=5cm]{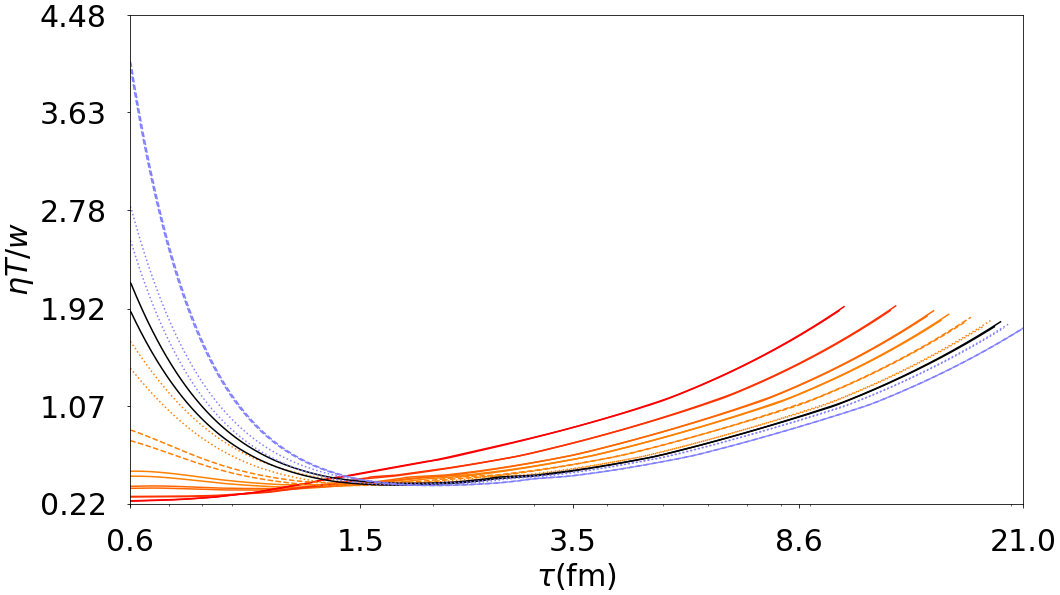}
    \caption{}
    \label{fig:eta}
    \end{subfigure}
     \begin{subfigure}{.48\textwidth}
     \includegraphics[width=\linewidth,height=5cm]{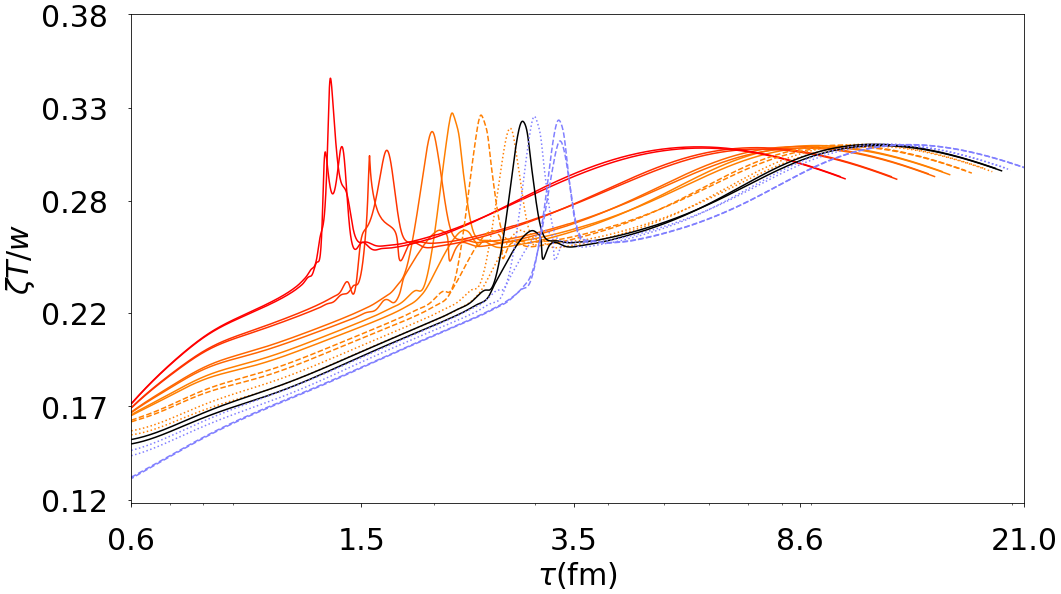}
     \caption{}
     \label{fig:zeta}
     \end{subfigure}
     \caption{Fig (a) shows the time evolution of $\eta T/w$ for hydrodynamic runs with different initial conditions. Fig (b) shows the time evolution of $\zeta T/w$ for hydrodynamic runs with different initial conditions having some critical point sensitivity.}
\end{figure}
The transition between the hadron gas phase and the QGP phase is matched using a $\tanh$ function in order to ensure a smooth matching.  Additionally, we ensure here that  the transition  line matches that in the equation of state (EOS). At $\mu_B=0$ we ensure that the shear viscosity to entropy density ratio has a minimum of $ 0.08$, which occurs at $T_{\eta/s,min}=196$ MeV, the $\mu_B$ dependence of the minimum $\eta T/w$ is determined by the change in $\eta T/w$ with $\mu_B$ within the excluded volume HRG model. The minimum of $\eta T/w$ converges to the transition line from the EOS at the critical point. Further work will extend this viscosity to the full 4 dimensional phase diagram of  $\left\{T,\mu_B,\mu_S,\mu_Q\right\}$ taking into account strangeness and electric charge chemical potentials.

For the bulk viscosity, we use the parameterization
 \begin{eqnarray}\label{eqn:zetanorm}
     \frac{\zeta T}{w} =\frac{9}{2\pi}\left(\frac{1}{3} - c_s^2\right)
 \end{eqnarray}
which has a maximum $\frac{\zeta T}{w}\sim 0.3$ similar to that found in certain hydrodynamic simulations \cite{Ryu:2015vwa}. The inclusion of dependence on $c_s^2$ allows the bulk viscosity to have some sensitivity to critical dynamics, as it is expected to drop considerably at the critical point \cite{Parotto:2018pwx} while the bulk viscosity is expected to diverge \cite{Monnai:2016kud}.

 In this work we use an equation of state that captures effects of criticality at finite baryon chemical potential. To this end, we use the most up to date equation of state reconstructed from Lattice QCD results, which are then mapped to a parameterized 3D Ising Model \cite{Parotto:2018pwx}. Here we place a critical point at $(T,\mu_B) = (143,350)$ MeV, with parameters $\alpha_1 = 3$, $\alpha_2 = 93$, $\omega = 1$, and $\rho = 2$, which was a default EOS setting from \cite{Parotto:2018pwx}. We note that the current version of this EOS is only reliable up to $\mu_B\sim 400$ MeV.  We, however, extend this out to $\mu_B \sim 600$ MeV in the high temperature region. That being said, we find limitations due to the maximum $\mu_B\sim 600$ MeV when studying the critical point. Finally, at this time we assume $\mu_S=\mu_Q=0$ but in future studies plan to incorporate the full 4D EOS in $\left\{T,\mu_B,\mu_S,\mu_Q\right\}$ from \cite{Noronha-Hostler:2019ayj} but we note that the 3D Ising critical point has not yet been implemented in the 4D EOS. 
 
 \section{Trajectories in the QCD Phase Diagram}
\begin{figure}
    \centering
    \includegraphics[width=\linewidth]{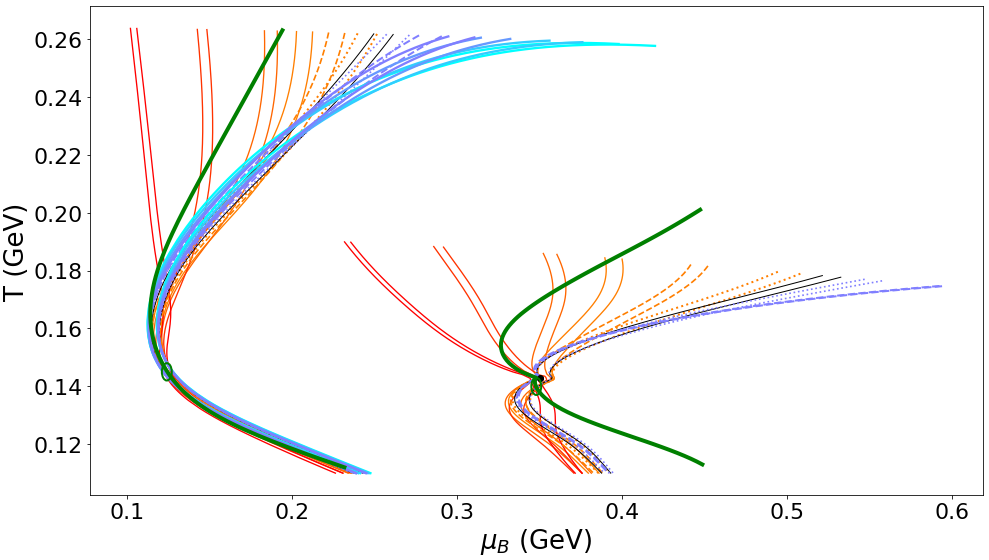}
    \caption{The $T-\mu_B$ trajectories of multiple hydrodynamic runs with different initial conditions, which all go through the same freeze-out region (green circle). The green line is the isentrope, for reference. For the isentrope away from the critical point, $S/N_B = 65$, and for the other $S/N_B = 21$.}
    \label{fig:traj}
\end{figure}
In Fig.\ \ref{fig:traj} we display the thermodynamic trajectories generated by hydrodynamics for a range of different initial conditions where the ratio between the dissipative stresses and the enthalpy
\begin{eqnarray}
    \chi &\equiv& \pi^{\eta}_{\eta}/(\epsilon+p)\\
    \Omega &\equiv& \Pi/(\epsilon+p)
\end{eqnarray}
are varied between $\pm 0.5$. The initial energy density for the trajectories away from the critical point is $7.0$ GeV/fm$^{3}$ and $1.5$ GeV/fm$^3$ for those that pass through the critical region. We place a constraint on the trajectories produced by imposing that all trajectories must pass through a freeze-out region defined along an isentrope close to where one would expect the system to freeze-out at given thermal fits \cite{Ejiri:2005uv,Schmid:2008sy,Bellwied:2016cpq,Noronha-Hostler:2019ayj,Motornenko:2019arp,Bellwied:2019pxh,Alba:2020jir}. The initial baryon density is varied until trajectories are found that pass through either side of the freeze-out region. The region was chosen sufficiently small to mimic the uncertainty in the freeze-out temperature and is represented in Fig.\ \ref{fig:traj} by a small green circle which trajectories pass through. If it were true that the system evolved isentropically, then one could uniquely determine the path traversed by the system given its freeze-out temperature and chemical potential. This path would correspond to the solid and thicker, green line (also shown). In this work, we argue that when considering viscous corrections, the unique correspondence between a freeze-out state and a trajectory in the QCD phase diagram no longer exists.
\par
\begin{figure}
    \centering
    \includegraphics[width=\linewidth]{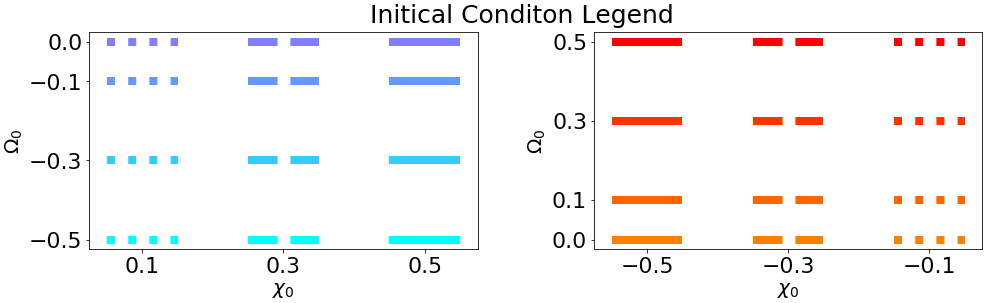}
    \caption{Reference for identifying which lines correspond to which initial conditions. Not shown here is the legend for $\chi_0 = \Omega_0 = 0$, which is represented by solid black curves in other figures. Note that not all combinations of $\chi_0$ and $\Omega_0$ are used, especially not in those trajectories that pass through the critical region.}
    \label{fig:Legend}
\end{figure}

 It can be seen in Eq.\ \ref{epDot} that shear and bulk out of equilibrium effects both contribute linearly for determining the dynamic behavior of $\epsilon$, albeit with opposite sign (see Table \ref{fig:Legend} for the different initial conditions used). Directly, one can notice that positive shear and negative bulk contributions work to slow down the rate at which energy density decreases. With this in mind, when looking at Fig.\ \ref{fig:traj}, it is consistent that those trajectories with the largest positive shear and negative bulk initial contributions start out the flattest in temperature and therefore must start at larger chemical potential in order to make it to the same freeze-out point as the other trajectories. We note that limitations in the EOS for producing data at large chemical potential ($>600$ MeV) prevented us from being able to explore the full range of initial positive shear and negative bulk, as was done for trajectories away from the critical point at low $\mu_B$.
\par
Again in Fig.\ \ref{fig:traj}, one can see a spread in initial $\mu_B$ of about $350$ MeV and a spread in initial temperature of about $10$ MeV. In order to better constrain the  trajectory path, it is important to be able to calculate the initial conditions for the system (which include the initial values for the dissipative stresses) precisely. Currently there exists some models for hydrodynamic initial conditions at both zero \cite{Gale:2012rq,Werner:1993uh,Kurkela:2018wud} and finite chemical potential \cite{Weil:2016zrk,Werner:1993uh,Shen:2017bsr}. However, the initial condition models at finite $\mu_B$ do not incorporate a fully initialized $T^{\mu\nu}$ such that we have no way to estimate how wide of fluctuations to anticipate in $\pi^{\mu\nu}$ and $\Pi$ on an event-by-event basis.  
 
 \section{Hydrodynamic Attractor}
\label{attract}

 In recent studies, it has been suggested that much of the success of relativistic viscous hydrodynamics in describing the system with such little knowledge of the initial conditions can be attributed to the existence of an attracting solution in the equations of motion of the viscous fields. In this work, we expand on this by studying a non-conformal hydrodynamic system with a finite chemical potential.
 \par
A natural question to ask when using more realistic transport coefficients that depend on temperature and chemical potential is whether or not the hydrodynamic attractor persists.
Typically, one would expect that when scaling the time evolution of these quantities by their respective relaxation times of $\tau_{\pi}$ and $\tau_{\Pi}$, these quantities would collapse onto a non-zero and non-trivial attractor and then eventually evolve towards zero (i.e., equilibrium). However, past work has primarily only focused on shear viscosity, which then only has one characteristic time scale.  Here we have two and a nontrivial equation of state.  In this work, we vary the initial values $\chi_0$ and $\Omega_0$ between $\pm 0.5$.

\begin{figure}
    \centering
    \begin{subfigure}{.48\textwidth}
    \includegraphics[width=\linewidth,height=5cm]{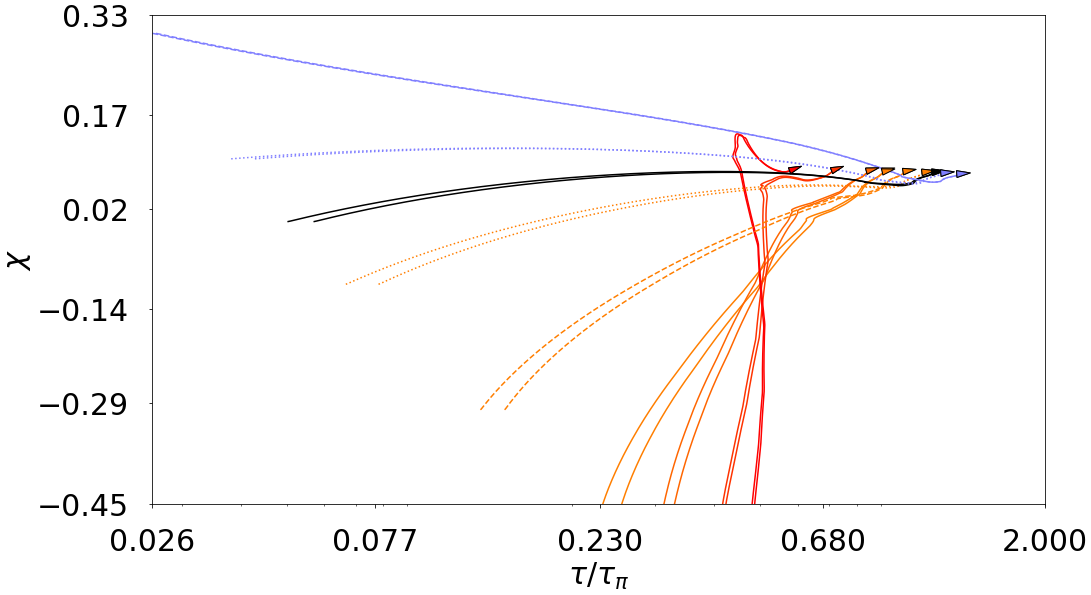}
    \caption{}
    \label{fig:chiSh}
    \end{subfigure}
    \begin{subfigure}{.48\textwidth}
    \includegraphics[width=\linewidth,height=5cm]{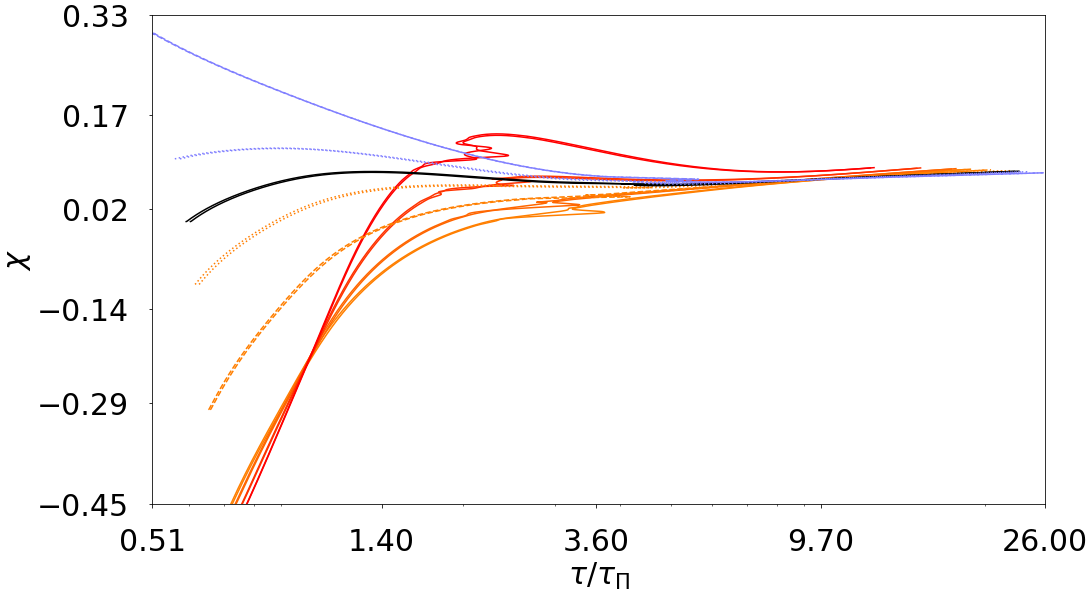}
    \caption{}
    \label{fig:chiBu}
    \end{subfigure}
    \caption{Fig. (a) shows time evolution of $\chi$ rescaled by its relaxation time. The end arrows point along the instantaneous directional derivative. Fig. (b) shows time evolution of $\chi$ rescaled by the bulk relaxation time.}
\end{figure}

\begin{figure}
    \centering
    \begin{subfigure}{.48\textwidth}
    \includegraphics[width=\linewidth,height=5cm]{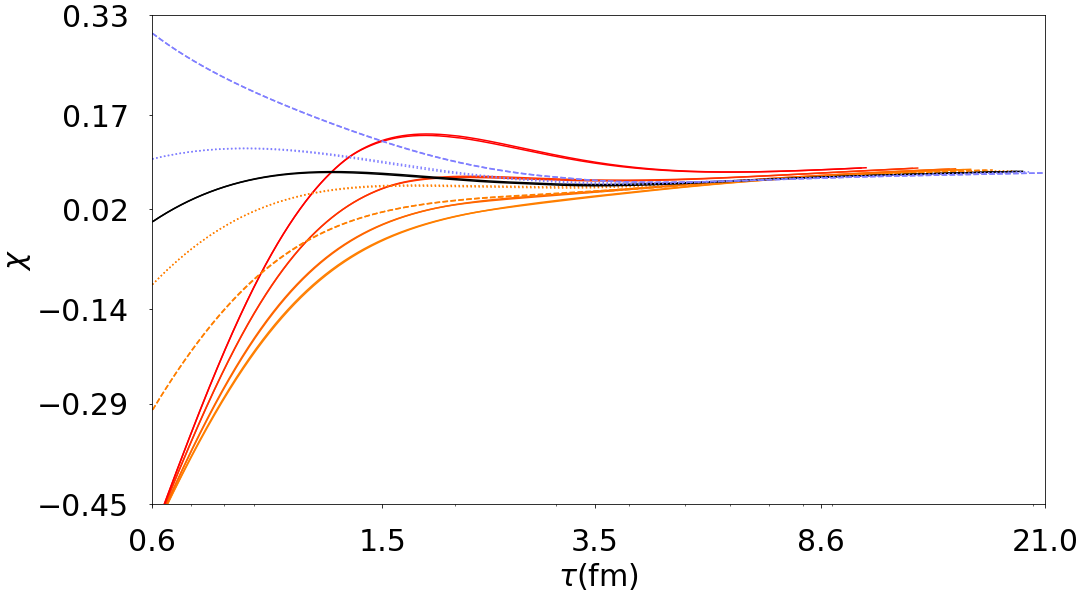}
    \caption{}
    \label{fig:chi}
    \end{subfigure}
    \begin{subfigure}{.48\textwidth}
    \includegraphics[width=\linewidth,height=5cm]{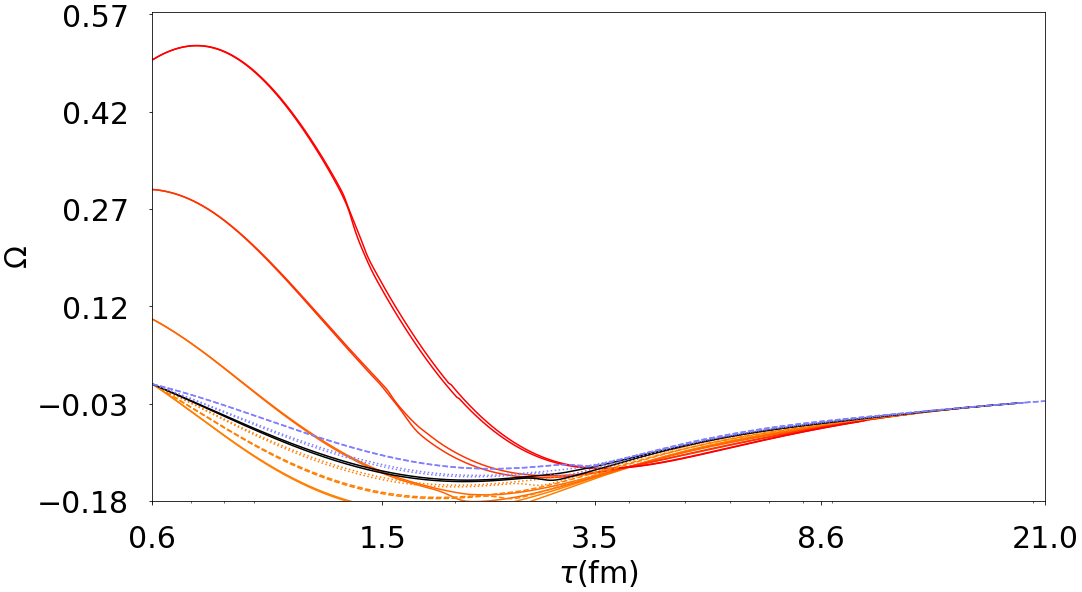}
    \caption{}
    \label{fig:omega}
    \end{subfigure}
    \caption{Fig (a) shows time evolution of $\chi$ without rescaling. Fig (b) shows time evolution of $\Omega$ without rescaling}
\end{figure}

In Figs.\ \ref{fig:chiSh}, \ref{fig:chiBu}, and \ref{fig:chi} one can see that the presence of an attracting solution is likely, even though the system never fully reaches equilibrium (since $\chi$ approaches a constant $\neq 0$ at late times). The added help of the arrows pointing in the direction of the $\chi$ trajectory at the final time in Fig.\ \ref{fig:chiSh} allows one to see that while the curves have not collapsed onto each other they do appear to be converging to a single line. At infinitely large times, it is likely that the $\chi$ trajectories under this rescaling would also collapse on to each other. Interesting to note, is the presence of weak attraction in the unscaled evolution, Fig.\ \ref{fig:chi}. There has not yet been much evidence to support the existence of an attractor in the unscaled time evolution. One should also note that the weak attraction seemingly present in Fig.\ \ref{fig:chiBu} may simply be due to the fact that $\tau_\Pi$ is of $\mathcal{O}(1)$.

\begin{figure}
    \centering
    \begin{subfigure}{.48\textwidth}
    \includegraphics[width=\linewidth,height=5cm]{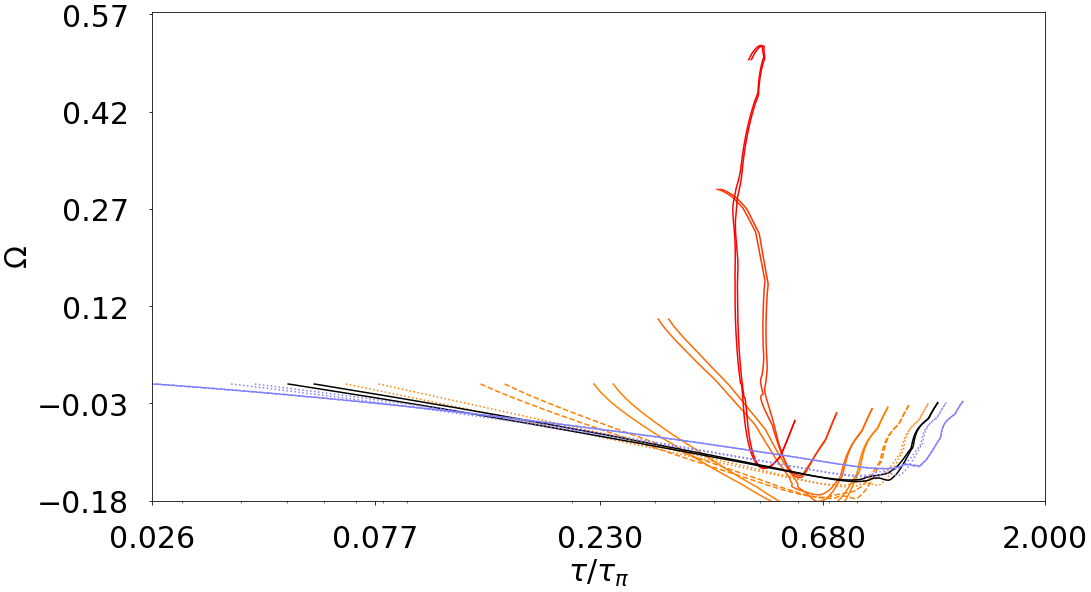}
    \caption{}
    \label{fig:omegaSh}
    \end{subfigure}
    \begin{subfigure}{.48\textwidth}
    \includegraphics[width=\linewidth,height=5cm]{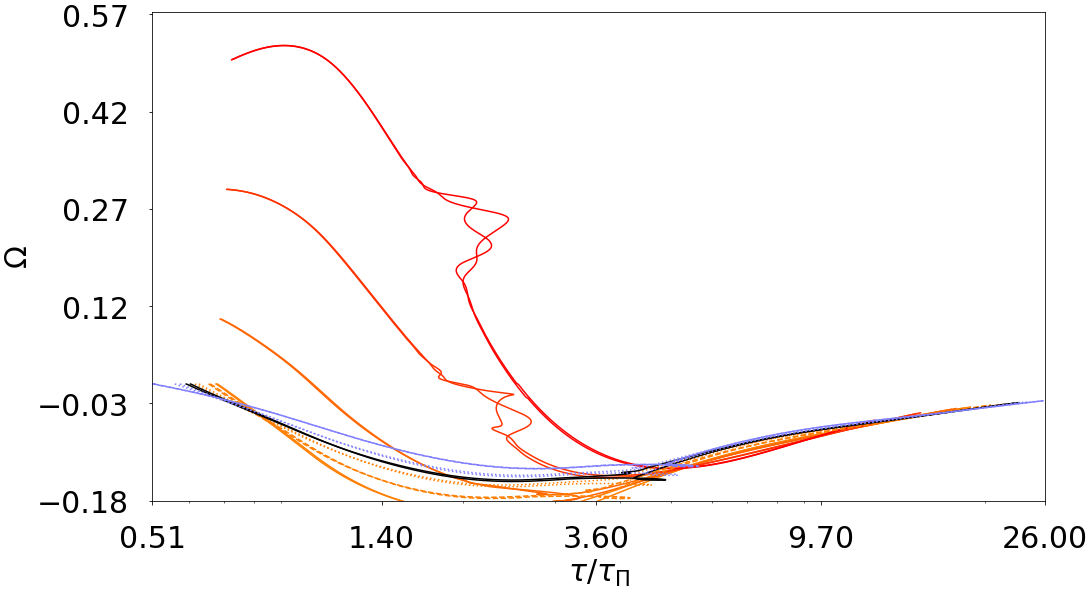}
    \caption{}
    \label{fig:omegaBu}
    \end{subfigure}
    \caption{Fig. (a) shows time evolution of $\Omega$ rescaled by the shear relaxation time. Fig (b) shows time evolution of $\Omega$ rescaled by its relaxation time.}
\end{figure}

\par
This explanation seems consistent with the fact that the attracting behavior for both $\chi$ and $\Omega$ in their unscaled evolution appears stronger than when their evolution is scaled by the bulk relaxation time, $\tau_\Pi$. This seems especially noticeable in comparison between Figs.\ \ref{fig:omegaBu} and \ref{fig:omega}. However, this notion goes against the intuition that the scaled time should be controlling the attracting behavior for its relevant quantity. Here we offer no solution to this slight paradox, but only point out its existence.
\par
Either way, Fig.\ \ref{fig:omega} shows a clear non-trivial and non-zero attractor for $\Omega$. Even as the system goes through the critical point, which the bulk viscosity is sensitive to via $c_s^2$, it is about an order of magnitude smaller than the shear viscosity. The implications are interesting given that the system exhibits memory effects of the initial state and never actually relaxes to equilibrium. While we are still missing critical fluctuations in this approach, it is interesting to see attractor-like behavior in the bulk viscosity near the critical point. Given that there is no reason to believe initial viscous effects in heavy-ion collisions at BES energies should be small, theoretical models should take far-from-equilibrium initial conditions into consideration.

\section{Conclusions}
Opening up of the new degree of freedom, $\mu_B$, in conjunction with exploring hydrodynamic non-equilibrium effects, lead to system dynamics that must be more thoroughly understood. In connection with the Beam Energy Scan program at RHIC, there are many important implications. For instance, given an initial $\epsilon$ and $\rho$, widely different trajectories may be seen throughout the phase diagram depending on how far-from-equilibrium the system begins. For these reasons, it is crucial to begin studying more deeply how to properly initialize the full $T_{\mu\nu}$ as input to hydrodynamics, compatible with event-by-event fluctuations, non-conformal system dynamics, and a finite chemical potential. Ongoing efforts such as \cite{Martinez:2019jbu} make it possible to study diffusive dynamics in the more heavily studied $\mu_B = 0$ regime by taking into consideration local fluctuations of the conserved charges. Moving forward, this seems to be a promising framework to begin asking the same kinds of questions asked here, but in a more realistic scenario.

\section*{Acknowledgements}
J.N.H. acknowledges support from the US-DOE Nuclear Science Grant No. DE-SC0019175, the Alfred P. Sloan Foundation, and the Illinois Campus Cluster, a computing resource that is operated by the Illinois Campus Cluster Program (ICCP) in conjunction with the National Center for Supercomputing Applications (NCSA) and which is supported by funds from the University of Illinois at Urbana-Champaign.

\section*{References}
\bibliography{BIG}

\end{document}